\documentclass[journal]{IEEE_JAIC}

\ifCLASSINFOpdf
   \usepackage[pdftex]{graphicx}
\else
\fi

\usepackage{xcolor}
\usepackage[utf8]{inputenc}
\usepackage {hyperref}

\begin{document}
\title{IoT Analytics and Blockchain}
\author{Abbas~Saleminezhadl, Manuel~Remmele, Ravikumar~Chaudhari, Rasha Kashef\\ Electrical, Computer, and Biomedical Engineering Department\\ Ryerson University \\ $\{$ abbas.saleminezhad, rchaudhari, manuel.remmele, rkashef $\}$@ryerson.ca}

\markboth{}
{Shell \MakeLowercase{\textit{}}}
{\maketitle}

\IEEEpeerreviewmaketitle

\begin{abstract}
The Internet of Things (IoT) is revolutionizing human life with the idea of interconnecting everyday used devices (Things) and making them ‘smart’. By establishing a communication network between devices, the IoT system aids in automating tasks and making them efficient and powerful. The sensors and the physical world, connected over a network, involve a massive amount of data. The data collection and sharing possess a critical threat of being stolen and manipulated over the network. These inadequate data security and privacy issues in IoT systems raise concerns about maintaining authentication of IoT data. Blockchain, a tempter-resistant ledger, has emerged as a viable alternative to provide security features. Blockchain technologies with decentralized structures can help resolve IoT structure issues and protect against a single point of failure. While providing robust security features, Blockchain also bears various critical challenges in the IoT environment to adapt. This paper presents a survey on state-of-the-art Blockchain technologies focusing on IoT applications. With Blockchain protocols and data structures, the IoT applications are outlined, along with possible advancements and modifications.

\end{abstract}


\section{Introduction}
The number of IoT devices will be approximately 29 billion by 2022 \cite{a1}. Every device generates and transfers data on the Internet. Considering the massive number of devices, it is easy to understand the crucial and continuous data production process. Different aspects of our daily life, including our cities, healthcare, transportation, energy systems, and homes, have been influenced actively by IoT technology. Various types of devices by the power of IoT can produce and transmit data across the network, which consist of laptops, computers, and phones but also sensors in the house, cars, traffic cameras, and common household appliances connected to the Internet. The market capacity of IoT that would be generated by 2020 is over 800 billion dollars revenue which opens a huge business future for different organizations \cite{a2}. On the other hand, the design and management of IoT will be more challenging because of its large scale and diversity.

Since the arrival of the blockchain (BC), certain types of issues related to IoT have been solved using this trustable, distributed ledger technology. The distributed and decentralized nature of BC is the reason for its popularity among the public \cite{a3}.  Traditionally, centralized servers like cloud servers oversee storing data of IoT systems for future usage. So, IoT users, for their privacy and securing their private data, must develop centralized servers. Despite the indisputable benefits that these centralized service providers present, they might face specific security issues. For instance, hacking of unencrypted data servers can cause the spreading of sensitive information \cite{a4}. Considering all these factors, IoT devices' management and data storage allowed industries to move towards decentralized architectures. Blockchain permits two devices to communicate and transfer resources, information, and data in a decentralized peer-to-peer network. As a gain of using Blockchain, it provides an infrastructure that minimizes the chance of any fraudulent entry because of its unique method for adding a piece of new information because the decision of adding new data will go through the network when many users instead of a single centralized unit approve the decision then it will take place. Therefore, the safety of blockchain-based IoT systems is guaranteed when attackers or hackers want to take control of a centralized server and get personal information. Taking advantage of blockchain, the cost of additional security monitoring of IoT servers can be diminished \cite{a5}. Blockchain technology storage systems used for IoT devices are in a way that data can be tracked easily in case of any contradiction or problem. The encryption methods used in blockchain ensure that if an intruder or adversary wants to infer the information illegally within a blockchain network, they will not be successful \cite{a6}.

In this paper we provide a comprehensive overview of the intersection of BC and IoT technologies. We demonstrate how BC can help to solve these issues. In addition, we illustrate blockchain applications in IoT within smart homes, smart cities, Supply chains, and Applications in Energy domains. This work also describes and summarizes the methods and other tools used in blockchain IoT. Moreover, we survey research trends and future works. This paper is then organized as : Section 2 discusses the background on IoT architecture and security analysis, and the Blockchain technology. Section 3 presents the Blockchain in the context of IoT. Section 4 discusses the blockchain tools, while the various applications are discussed in section 5. The paper is concluded with future directions in section 6.

\section{Background}
Blockchain can help mitigate the security and scalability concerns of the IoT environment.\cite{r12} But first, we should understand the architecture of the IoT system in order to understand its security issues. Then, we will discuss the Blockchain and its features(methods) that can help with these issues. In the following section, we will discuss the IoT architecture, security analysis and blockchain introduction.

\subsection{IoT architecture overview}
\begin{figure}[h]
\includegraphics[width=8cm]{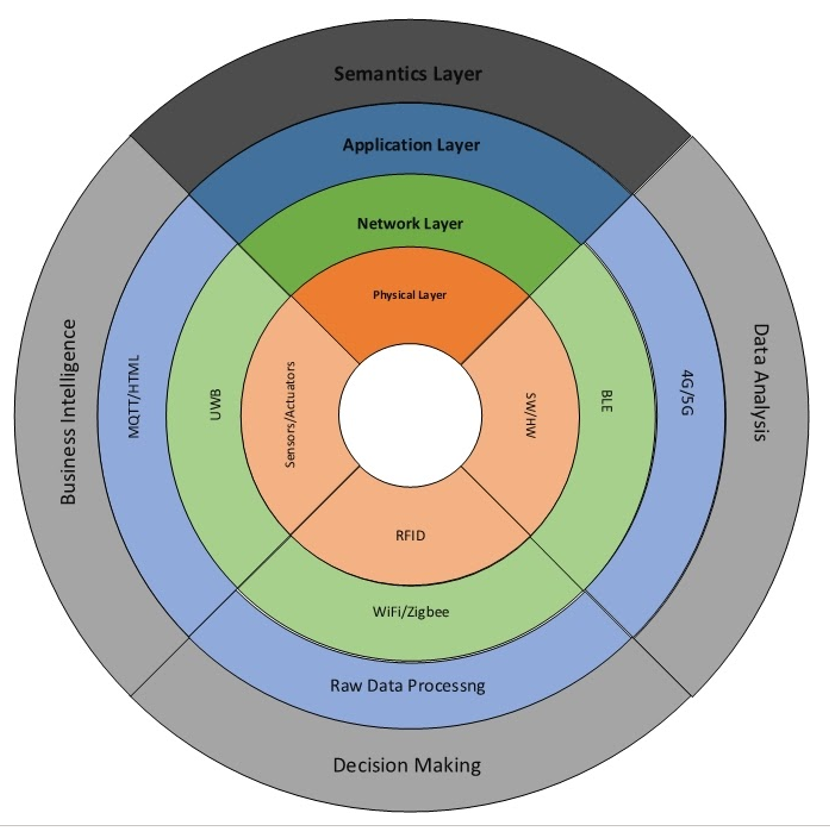}
\caption{IoT Architecture}
\label{fig}
\end{figure}

The rapid growth in the popularity of IoT systems has caused many organizations to spend a large amount of funding in research to find an IoT solution. Therefore, the industry focused on producing the solutions often builds all components of the stack, from hardware to cloud services. This approach has resulted in the inconsistency of IoT architecture \cite{r1}. From several ideas and literature reviews, we can conclude an architecture that can be seen as the fundamental architecture of an IoT system. A layered IoT stack from paper \cite{r2} is represented in fig ~\ref{fig}. The Physical or Perception layer is composed of various devices and sensors, gathering a continuous data stream. These devices are often deployed as a star, clustered tree, or mesh topology in the network. The devices or the “Things” are connected to a centralized gateway with various protocols such as Zigbee, Bluetooth Low Energy, LwM2M and Wi-Fi. The third layer is the Application layer, where the gateways are connected to servers via LTE, Optical Fiber Cable, 5G etc. This layer provides analytical services to the end-user to create a perception of the data collected from the sensors. Especially in the third layer, challenges such as data privacy, processing, handling, and storage are encountered. For example, health sector governing bodies of many countries prevent information sharing of patients to any other parties without consent. The last layer is the Semantics layer which can be deemed as a data analysis and business intelligence layer \cite{r2}.

The Physical or Perception layer is composed of various devices and sensors, gathering a continuous data stream. These devices are often deployed as a star, clustered tree, or mesh topology in the network. The devices or the “Things” are connected to a centralized gateway with various protocols such as Zigbee, Bluetooth Low Energy, LwM2M and Wi-Fi. The third layer is the Application layer, where the gateways are connected to servers via LTE, Optical Fiber Cable, 5G etc. This layer provides analytical services to the end-user to create a perception of the data collected from the sensors. Especially in the third layer, challenges such as data privacy, processing, handling, and storage are encountered. For example, health sector governing bodies of many countries prevent information sharing of patients to any other parties without consent. The last layer is the Semantics layer which can be deemed as a data analysis and business intelligence layer.\cite{r2}

\subsection{IoT Security analysis}
According to \cite{r3}, All IoT devices are vulnerable to fragile security provisions such as weak passwords, lack of encryption in the network and poor device service management. This makes IoT devices an easy target to attack and steal data. Also, IoT devices in the network are often installed in unattended areas with low surveillance. These are some of a few fundamental issues with IoT system security. Since devices are connected to the internet, they are prone to security attacks on each layer of the IoT system. In the following section, we have tried to summarize these attacks based on papers \cite{r2}, \cite{r3} and \cite{r4}.

\subsubsection{Physical Layer Attacks}
These types of attacks are focused on the hardware part of the system, and generally, the attacker needs to be close to the device to carry these attacks. Node Tampering, Node Jamming in WSNs, Fake Node Injection, Physical Damage and RF interference are some of the physical layers of well known attacks. Attacks on nodes are usually carried by replacing the entire node with a fake node. Then attackers can control the data flow to inject malicious data into the network and steal existing information. These attacks can be harmful for hardware lifetime and their functionalities

\subsubsection{Network Layer Attacks}
When we think of a cyber-attack on an IoT device, network attacks come first in our minds. Based on the paper's review of \cite{r2}, \cite{r3}, and \cite{r4} some of the common attacks are Sybil attack, Denial of Service and TCP/IP. In the Sybil attack, a malicious node is presented to claim a large number of nodes. These attacks are especially dangerous in the voting system of the sensor network where the Sybil node votes more than just once and manipulates the results. TCP/IP protocol attackers use IP source address manipulation to spoof authentication and then control the network in the Application layer.

\subsubsection{OS or Firmware Attacks}
To guarantee the security and privacy of the IoT devices, we must have Software, application, configuration, and OS integrity. Recently, the world faced an attack called “Mirai”, which created a botnet by attacking thousands of IoT devices. It exploited the software weakness to hack through thousands of devices and created a DDoS attack on DNS. It basically created thousands of lookup requests to DNS, so the DNS system collapsed and stopped responding. This creates a single point failure which can shut the whole system. Apart from this, software attacks can be phishing, virus, malware, spyware, and malicious scripts. Lack of security measures mechanisms such as anti-virus across the IoT network leads to attacks on software and applications.

\subsubsection{Sensor data Attacks}
Attackers usually target network protocols to get access to the data of network devices. As mentioned in \cite{r6}, IoT networks use ad hoc protocols, i.e., information is transmitted hop-by-hop until it reaches the destination. This type of protocol creates opportunities for the attacker to forward the tampered data into the system and manipulate the behavior of a system. For example, if attackers target the traffic data of a GPS, then they can create traffic congestion by using false traffic information. Authentication algorithms have been created; however, they can be easily exploited by normalized tampered data with original data.

\subsection{Blockchain Technology}
Satoshi Nakamoto introduced the concept of blockchain. The blockchain is a distributed system running over a peer-to-peer network and does not need a central unit for authority, so verification of 3rd party units is no longer needed. At first, blockchain was introduced for cryptocurrency systems for its unique way of recording transactions; However, various fields can benefit from the evolution of blockchain.   	                                                                                                                                           

The blockchain is designed for times that participants in the network do not trust each other so it goes to record information with respect to it. So, verification and accountability will be provided by this public and auditable ledger \cite{a7}. Every blockchain consists of some blocks from the genesis block to the current block. Because every block contains a hash of the previous block it makes sense to see it as a chain of blocks. The first block in a blockchain is the genesis block that is always hard coded into the software. All transactions are visible in the blockchain. When there is a new transaction, this transaction is going to be written down in the local ledger by a group of  volunteer recorders  in that local ledger. After some time, a selected ledger from these recorders will be verified which block contains a set of transactions and will be connected to the public one \cite{a8}. In blockchain systems, miner’s duty is to record these transactions and this process is called mining. They designed Mining to be resource-intensive and hard purposefully. Every individual block should have proof of work (PoW) \cite{a9} this is the way to evaluate the validity of a block in the blockchain. Once a miner receives a block The PoW will be verified by other miners. By this method, secure and tamper-resistant consensus that is the goal of mining will be satisfied. It takes time to spread a transaction through all the nodes in the network, and this time delay is used to make sure that all the nodes in the network have verified all the transactions. The consensus algorithm is an algorithm for selecting a block to connect to the main public ledger. In distributed systems, Consensus is a primary problem that requires more than one agent to decide on a given value needed for computational purposes. It can be assumed that some of these agents are unreliable, therefore the consensus process needs to be creditable.

 These selected blocks form a public ledger that is called the blockchain \cite{a10}. Blockchains can utilize different consensus algorithms and proof of work (PoW) is one of them \cite{a9}. PoW of every block assures that there is a certain amount of difficulty to generate a new block in the network. new blocks are going to connect to the blockchain If there is consensus to accept that new block, so all the miners should start mining using that block considering it as a reference. A digital signature protects this whole procedure; in a way that each transaction should be digitally signed into the network using the private key of the sender. This will guarantee the validity and integrity of transactions. Hash computations are there to assure the integrity of the whole blockchain system. Coins are assigned as a reward to miners who participate in successfully appending a new block to the ledger \cite{a10}.

\section{Blockchain in the context of IoT}
Blockchain can provide a decentralized, safe, and secured data sharing ledger in which data can easily be traced, so it has the capability to efficiently revolutionize the functioning of a large number of IoT services and applications. It can improve diverse IoT systems such as smart homes, smart cars, energy, smart city, healthcare systems, etc. considering features of blockchain, it can be stated that IoT can benefit from blockchain concepts significantly. Here we mention the main progress applied by integrating blockchain and IoT.

The first feature is Decentralization. Main centralized security issues in IoT applications are overcome by the decentralized nature of blockchain,preventing security issues like a single point of failure to guarantee the responsivity and promptness of IoT services or prevent situations where some companies control the storage and processing of data of numerous people. Furthermore, it can help IoT devices with reliable governance and management, and it can track IoT devices for their entire life cycle. Additionally, in blockchain, all nodes must have a copy of the data, and therefore, data is immutable and reliable. Moreover, it can offer privacy to the network in which transactions can be approved in a reliable network, so nodes are unknown, and it can be assumed that information alongside their identities can be preserved \cite{a11}].
The security and immutability that blockchain brings to IoT are remarkable. By utilizing cryptographic encryption, all transactional information of IoT applications will be protected \cite{a12}. Transactions stand for the data transmission across various devices in blockchain-based IoT systems, and data will be secured due to transactional protocols. 160-bit address space that blockchain offers as a hashed public key to assign a huge number of addresses that can be shown secure and unique to be allocated in IoT devices. Moreover, a 160-bit address compared to another addressing scheme like IPv6 that offers 128-bit for addressing, is beneficial and can satisfy the huge IoT demand for future applications. Additionally, all the transactions taking place in the network can be tracked down and cannot be changed to assure data immutability and reliability. Blockchain offers smart contracts, as in Ethereum. Privileges and access controls across nodes in the network can be defined as hard code rules. They can provide decentralized authentication logics that are less complex and hardcoded into rules to authorize IoT devices effectively. It also maintains data security when configuring conditions and criteria under which certain nodes can access specific data. The trust of participants can be guaranteed by the immutable nature of blockchain-based IoT systems, as they can track down and confirm any transaction without any risk of tampering \cite{a13}. The third property of blockchain is Identity. In blockchain-based IoT systems, for every connected device, information can be traced easily as unique identifiers. Authorized and trustable identity control of connected IoT devices can be provided by this identity feature accompanied by their complex features and relationships. Every stage in an IoT device life cycle from the producer, supplier, and consumer can be tracked \cite{a14}.

\section{Methodes}
To enable Blockchain for IoT applications additional tools are required. Therefore, very valuable are the InterPlanetary File System (IPFS), an emergent technology, and smart contracts that are already widely used. Another approach for distributed databases is Bigchain DB.  To implement the blockchain itself, a lot of research papers are taking advantage of Hyperledger due to its big variety and modular appearance.

\subsection{IPFS}
For current IoT applications, the data is stored in centralized servers. The reliability, privacy, and
interoperability of these data can be increased by storing it decentralized. One IoT framework for decentralized data storage based on the InterPlanetary File System (IPFS) as described in \cite{ipfs}.

The IPFS is a System to share content like HTTP or FTP. But there are a few main differences to the common systems:
\begin{itemize}
\item Addressing content: In comparison to the Internet as we know it, in the IPFS you are not addressing the location of the content but the content directly. So, to access a file the hash of this file is used. Then a ping for this file is sent. Because the hash is used to access the file it is also secure against attacks on the consistency of a file. These hashes are forming a Merkle Directed Acyclic Graph (DAG). 
\item Decentralization: As the content and not the localization of the file is relevant, the localization can change without causing problems. Whenever a file is downloaded from a new node, it can be accessed from this node as well. This leads to a lot of advantages, such as security against attacks on the availability or lower bandwidth. 
\item Size of an object: The maximum size of a file is 256 kb. To be able to store bigger files they are split into multiple objects and an additional object needs to be created that links all objects.
\item Usage as a file system: The data architecture allows it to use IPFS as a file system. In this case, an object includes the hashes of multiple other objects to link them as a folder in a common system includes the location of the objects that are stored inside it. This architecture leads to immortality as it is not possible to delete a hash out of an object without changing its hash. To be able to change an object nevertheless, versioning is supported by IPFS.
\end{itemize}

The PFS can be used to store IoT data decentralized. It erases the need for a central node and the disadvantages in security and stability coming with a centralized system. As Blockchains have issues with gigantic amounts of data appearing in more IoT applications nowadays, IPFS enables the blockchain to handle this problem. This leads to a combination of widely used, well approved and known blockchains and IPFS to store the data. In \cite{ipfs2} an architecture was built to provide IoT data privacy via blockchains and IPFS. Access control was controlled by smart contracts.
 
\subsection{BigchainDB} 
The BigchainDB is another decentralized database through an extensible blockchain described in \cite{bcdb}. While combining the distributed databases with blockchain technology, we integrate the advantages of both. As BigchainDB is a modern distributed database, it has all the features as linear scalability, availability, fault tolerance and final consistency. Simultaneously, it features blockchain characteristics, such as decentralized control, non-tamper, Issuing and trading assets directly and autonomy. Paper \cite{bcdb} as well shows an improvement in time needed to query a transaction in comparison with HadhoopDB and Hive.

\subsection{Smart contracts}
Smart contracts are programs that get executed when a particularly relevant event is happening that was defined in this contract. The primary use case for this contract is to act like a normal contract to transfer money within a smart ecosystem—the best-known blockchain with smart contract functionality Ethereum. But the financial aspect of the IoT is not the only scenario for using smart contracts. When it comes to data, smart contracts can be helpful to control the communication access between IoT nodes.
 
\subsection{Hyperledger Sawtooth}
In 2015, Hyperledger was created as an open-source blockchain platform by the Linux foundation. Compared to other distributed ledger technologies (DLT) as Ethereum or R3 Corda, Hyperledger is broadly based with its immense variety in frameworks. Its modular appearance makes it easy to implement different applications such as smart contracts. Unlike other DLTs, Hyperledger doesn’t have its currency; however, coins can be implemented easily. In connection with IoT, two big frameworks are worth mentioning are:
\begin{itemize}
\item Hyperledger Fabric: As mentioned in \cite{eth_ipfs} Hyperledger Fabric is one of the promising DLTs to implement easy usable big data storage in blockchain.
\item Hyperledger Sawtooth: In \cite{saw_ipfs} Sawthoot was used to build an IoT network that took advantage of IPFS.
\end{itemize}

\section{Applications}
\subsection{Smart home}A Smart Home is a concept that utilizes IoT that can be used in every device that can generate and transfer data to increase resident comfort and reduce operation costs by efficiently managing resources. On-location controllers capable of gathering and processing data are provided for these smart homes, controlled through a device to personalize the environmental adjustments. Smart Home devices are increasing day by day. The financial potential of this data is vast because data generated from smart homes can be used in various applications. For instance, the healthcare industry may unravel severe health issues by analyzing data collected from smart home IoT-enabled devices to offer personalized healthcare services. Smart-home devices and IoT appliances are quickly becoming ordinary, but there are issues about security. Blockchain technology can provide a direct and obvious method, allowing you to control everything securely from a simple on/off switch or even more complex connected devices capable of more access and controls for things in the home \cite{a15}. Minors in the smart home are responsible for handling all data transmission inside and outside the home, always online and high-resource devices. Communications are controlled and audited by a private and safe blockchain preserved by a minor. A blockchain-based smart home framework should guarantee its safety by analyzing its security concerning the fundamental security aspects of confidentiality, integrity, and availability. Confidentiality guarantees that only authorized users can access information. Integrity guarantees that the sent data without any Interference is received at the destination, and availability makes sure that every service or data is accessible and attainable to the user whenever is needed \cite{a16}. Many devices in the home capable of connecting to a network are low energy and lightweight; therefore, most of their available energy must be devoted to computations and executing core application functionality. The power that remains for security and privacy is limited and made affordable of these tasks quite challenging. Energy consumption and processing overhead of Traditional security methods tend to be expensive for IoT. IoT in smart homes requires a distributed, delicate, scalable security and privacy system. The distributed, secure, and private nature of blockchain technology has the potential to overcome these challenges.

Integrating blockchain with smart home IoT is not simple. It involves several crucial challenges such as high resource demand for solving the POW, long latency for transaction confirmation, and low scalability resulting from broadcasting transactions and blocks to the whole network. In a study case, Dorri \cite{a17} proposed a new framework of blockchain by removing the concept of POW and the concept of coins as rewards. For making his framework more suitable for the specific requirement of IoT, His framework is based on a hierarchical structure and a distributed trust to maintain blockchain security and privacy. The design consists of three main components: smart home, overlay, and cloud storage. A miner centrally manages all Smart devices located inside the smart home. Smart homes consist of elements like an overlay network, Service Providers, cloud storage, laptops, smartphones, or personal computers. The overlay network is like a peer-to-peer network that provides distributed features to the framework. Cloud storage is responsible for storing and sharing data of smart home devices. A local and private blockchain offers secure access control to the IoT devices and their data. Also, the blockchain creates permanent time-ordered records of transactions connected to other framework components for giving specific services. Symmetric encryption offers to achieve lightweight security for smart home devices. It can be stated that common security frameworks are not necessarily optimized and suitable for IoT due to processing burden and energy consumption. Here, using a smart home as a case study is the idea to illustrate the challenge and solution for blockchain application in IoT.

\subsection{Smart cities}
Within all the data that must be handled in Smart cities comes a big need for security. Blockchain can solve this in combination with other tools, as the following proof-of-work papers show. In \cite{eth_ipfs} a car accident in a smart city is simulated. Smart contracts and highway sensor data are stored on an IPFS implemented in Ethereum. When the sensors recognize an accident, several smart contracts stored on IPFS are executed. This proof of work paper concludes that there are several issues on IPFS in combination with blockchain caused by the fact that it is a new field of study. These issues are, e.g. lack of support by blockchain platforms or non-existing standards. A similar scenario was simulated in \cite{saw_ipfs}. In contrast, Hyperledger Sawtooth was used. Furthermore, the Blockchain and the IPFS were implemented in separate networks. Thus, a private IPFS could be used to increase privacy. Because of the limited transaction validation rate of Hyperledger Sawtooth, the blockchain was limited in validation nodes and transaction rate.

\subsection{Applications in Energy Domain}
During our lifetime, we consume energy from a variety of sources. Each part of our daily tasks would not be possible without energy. As time passes, these energy sources are becoming more and more scarce around the globe.\cite{r13} There was no better time than now to start finding more ways to save these highly draining energy sources. Traditionally, either government or private firms control the generation and distribution of electrical power. However, change has been noted in past years with solar panels and residential energy storage proliferation.\cite{r7} The global market of Solar panels was nearly \$52.5 billion in 2018 and is expected to grow further to \$223.3 billion by 2026. Similarly, residential storage capacity will increase to 3.7 GW by 2025.\cite{r7} This will change the trend of centralized distribution to decentralized distribution of electrical power. This new approach also comes with its issue, which is extensively described in \cite{r8} and \cite{r11}. Blockchain technologies have received enormous attention to provide immutable, distributed transaction ledgers and transform nearly every type of global business. Consequently, researchers have also successfully found ways to integrate blockchain in smart grid applications. Among those,  we will discuss a few interesting applications of blockchain in electrical power distribution and consumption.
 
\subsubsection{Decentralized Energy resource management using Ethereum blockchain}
In the decentralized model of electrical power systems, synchronized information exchange is critical. Power and load level information should be declared to surrounding nodes to gain balance between all grids.\cite{r8} An information leak can be fatal, and it can cause the system to shut down. External and internal actors can initiate data leaks to exploit the system for personal gain. A research team from the USA experimented on ten college campus buildings. More details are available in \cite{r8}. They created a simulated environment of solar panel grids of 10 buildings and checked their performance based on season, time of the day, and factors like clouds. They also created a smart contract application on Ethereum Virtual Machine via a DApp. This experiment aims to create a synchronized and secure eco-system for the microgrid around the ten buildings with the help of Blockchain technology.

Micro-grid systems face similar authentication and identity management issues as other IoT applications. The blockchain application used in this experiment faced an identity issue with smart contracts. Smart contracts provided identifiers, but overhead work to create a new contract to claim information was too high. To Mitigate this issue, they proposed a solution with a new architecture in which they required a unique identifier for the device or user, a way to authenticate this identifier, a way to make a claim and a way to authenticate this claim. The smart meter will communicate with Ethereum smart contracts over an untrusted network such as the Internet. The smart meter has been equipped with authentication gateway software. On the other hand, A node.js server will be set up with an Ethereum smart contract to validate this identity. To make this work, when a smart meter attempts to connect to Ethereum smart contract, a first packet authentication token will be sent with identity information. This token will through an untrusted network to node.js and be verified for further actions. This approach is very simplistic for such an application. The token created must go through the internet to reach Ethereum smart contract, which can be traced to get the identity. Moreover, over ten buildings on a low scale, the experiment was performed to identify the more significant issue as the system expanded. We should take this approach to the next level with a small area with multiple buildings to receive more information and solutions. Power efficiency was also affected after the implementation of Ethereum.

\subsubsection{Blockchain for renewable energy credits and carbon credits}
The renewable industries have incentives based on clean energy and reduced carbon usage to provide access to clean energy for end-users. Renewable energy credits (RECs) are based on one megawatt-hour (MWh) clean energy produced and one metric ton of CO2 avoided. The RECs can be purchased by end-users or businesses to claim the credit worth of energy that comes to them is from renewable sources. This great incentive is in place; however, it has lacked account and management procedures. The credits were stored manually in an Excel spreadsheet. This procedure not only was slow but also could fall into danger of double counting. The credits are often provided months later from the actual power generation\cite{r10}. The authors mentioned blockchain technology, especially the smart contract to automate and secure the entire procedure. This will make the accounting and management procedure secure and efficient. Clean Energy Blockchain Network (CEBN) and Silicon Valley Power formed a partnership to include blockchain technology in the Low Carbon Fuel Standards (LCFS) program. With this program, Credits are given to clean fuel providers, especially to Electric vehicle charging station owners. Then these credits can be sold to conventional fuel producers to compensate for their carbon emissions. Blockchain technology can verify clean energy produced with affiliation with Energy Engineers- Certified Energy auditor. The verification process aids RECs in blockchain technology from pre-qualified energy assets. This verification process saves time, cost, and resources.

\subsubsection{Secure Electric Vehicles charging in Smart Community}
Renewable energy sources and Electric Vehicles have been in great attention due to their promise of reducing fossil fuel use and cutting gas emissions to very low. Smart Community is an essential part of the Internet of Energy, which allows energy generation and distribution with a smart grid architecture. However, it is hard to ensure a secure schedule of charging behaviour of EVs in the untrusted energy market. In  \cite{r10}, An algorithm is proposed with the use of smart contracts from blockchain technology to secure the charging stations. First, they launched a novel permissioned blockchain system in the smart community. The number of pre-selected Electric Vehicles can access and publicly share the transaction with blockchain technology without interference from the trusted intermediary. After that, a consensus-based algorithm (delegated Byzantine fault tolerance) is presented to reach consensus in an effective means. Finally, the controlling operator evaluates and designs the best possible contracts to meet Electric Vehicles energy requirements. They have presented a performance evaluation where they showed performance based on three different schemes. In a flat rate scheme, the price was fixed regardless of the Electric vehicle type. With the increase of the Environment Coefficient($\omega$), price did not change; thus, the scheme worked poorly. For a two-way tariff scheme, price and demands are linearly related. Therefore, it shows slow growth in utilities. The third scheme, which the authors proposed, showed a great increase in the growth of utilities with the rise of $\omega$. This was achieved by designing optimal contracts.

\section{Conclusion and Future work }
This paper provided an overview of the IoT architecture, the security in IoT, and the state-of-the-art Blockchain. We have introduced Blockchain in the context of IoT added by some additional methods used concerning Blockchain and IoT. Finally, applications were pointed out, such as smart homes, smart cities, and applications in the energy Domain. Security challenges in IoT systems should be the priority to increase the acceptance of IoT applications among consumers. Blockchain technologies such as proof-of-work, smart contracts and private Blockchain are extensively used in the IoT industries for various applications. Our observation while conducting the survey is that Blockchain has the capacity to alleviate the IoT issues of scalability, network, security, and heterogeneous data. The Blockchain technology design would also adapt to properties of decentralized IoT networks, inherent partitioning, and various topologies. We also observed that implementation of Blockchain to IoT applications itself is a challenge and requires the core knowledge of the application. In a few applications, such as electrical microgrids and smart homes, Blockchain technology was adding a significant delay to the extent that researchers had to adopt a new way to reduce the delay. Therefore, Blockchain technology requires future research on optimizing faster transactions between two nodes or users and nodes. It was also shown that security and privacy blockchains are essential for the future of IoT.

\bibliographystyle{IEEEtran}
\bibliography{IEEEexample}

\end{document}